\documentclass[prd, aps, superscriptaddress, preprintnumbers, twocolumn, floatfix, nofootinbib]{revtex4}
\pdfoutput=1

\usepackage{amsfonts}
\usepackage{amsmath}
\usepackage{amssymb}
\usepackage{bm}
\usepackage{dcolumn}
\usepackage{graphicx}   
\usepackage[latin1]{inputenc}
\usepackage{latexsym}
\usepackage{rotating}
\usepackage{hyperref}
\usepackage{graphicx}
\usepackage{color}

\newcommand\be{\begin{equation}}
\newcommand\ba{\begin{eqnarray}}
\newcommand\ee{\end{equation}}
\newcommand\ea{\end{eqnarray}}

\begin{document}

\title{Clock Fields and Logarithmic Decay of Dark Energy}

\author{Robert Brandenberger}
\email{rhb@physics.mcgill.ca}
\affiliation{Department of Physics, McGill University, Montr\'{e}al, QC, H3A 2T8, Canada}

\author{Vincent Comeau}
\email{vincent.comeau@mail.mcgill.ca}
\affiliation{Department of Physics, McGill University, Montr\'{e}al, QC, H3A 2T8, Canada}

\author{Leonardo Fossati}
\email{lfossati@student.ethz.ch}
\affiliation{Physics Department , ETH Z\"urich, CH-8093 Z\"urich, Switzerland}

\author{Lavinia Heisenberg}
\email{lavinia.heisenberg@phys.ethz.ch}
\affiliation{Institute for Theoretical Physics, Heidelberg University, Philosophenweg 16, 69120 Heidelberg, Germany}
\affiliation{Institute for Theoretical Physics, ETH Z\"urich, Wolfgang-Pauli-Strasse 27, CH-8093,  Z\"urich, Switzerland}

\date{\today}

\begin{abstract}

We investigate the physical measurability of the infrared instability of a de Sitter phase in the formalism recently proposed by Kitamoto et al. \cite{Kitamoto}. We find that the logarithmic decay of the effective cosmological constant is only measurable if an additional clock field is introduced. 

\end{abstract}

\pacs{98.80.Cq}
\maketitle

\section{Introduction} 
\label{sec:intro}

De Sitter space plays an important role in the current paradigm of cosmology. On one hand, a period of accelerated expansion yields an interesting model of early universe cosmology, the inflationary scenario \cite{Guth}. On the other hand, current observations indicate that we are entering a phase of late time de Sitter-like expansion, the {\it Dark Energy} phase \cite{DE}. De Sitter space is also of special interest in General Relativity since it is a space with maximal symmetry (see e.g. \cite{dSRev} for a review of de Sitter space).

Various arguments, however, indicate that de Sitter space is plagued by an infrared instability which is already manifest at a semiclassical level in which one quantizes the fluctuations about the classical de Sitter background solution. Pioneering work on the instability of de Sitter space is due to Polyakov \cite{Polyakov}. Mottola and collaborators explored the instability due to fluctuations of the conformal mode \cite{Mottola}, Woodard and Tsamis \cite{WT} analyzed the instability arising from the back-reaction of gravitational waves, and the instability due to the back-reaction of scalar cosmological fluctuations was explored in \cite{ABM} (see also \cite{RHBBRrev} for a review) \footnote{Note, however, that there are also arguments claiming that de Sitter space is stable \cite{Marolf}.}. 

Under the assumption that superstring theory is the correct approach to unify all forces of Nature at the quantum level, there are arguments \cite{Banks, Dvali, swamp} (see also \cite{swamprevs} for reviews) which indicate that it is not possible to obtain stable de Sitter space, and only spaces with negative semi-definite cosmological constant are realizable. These arguments severely constrain canonical single field inflation models, they rule out a bare cosmological constant as an explanation for Dark Energy, and also yield constraints on quintessence models of Dark Energy \cite{Lavinia}. Recently, the {\it Trans-Planckian Censorship Conjecture} (TCC) has been put forwards \cite{TCC1} which also prohibits stable de Sitter space and leads to severe constraints on cosmological inflation \cite{TCC2} (see also \cite{TCC3} for some refinements).  The TCC can be viewed  \cite{TCC-Rev} as a generalization of Penrose's Cosmic Censorship Hypothesis \cite{Penrose} which prohibits the existence of black holes with charge greater than the mass \footnote{The TCC ccndition can also be obtained by demanding that the entanglement entropy between sub- and super-Hubble modes does not exceed the radiation entropy after inflation \cite{Omar}.}.

In this paper we shall focus on a recent approach to the infrared instability of de Sitter space developed by Kitamoto and collaborators \cite{Kitamoto} (see also \cite{Kitamoto2}) who find a logarithmic decay in the value of the effective Hubble constant as a function of time. We demonstrate that without a clock field, the effect computed in \cite{Kitamoto} is not locally measurable. However, once we introduce a clock field into the system, the effect becomes physically measurable and the ampitude and sign of the effect are as predicted in \cite{Kitamoto}.

In the following section we review the mechanism by which the back-reaction of cosmological perturbations can lead to an instability of a de Sitter phase. Section 3 is a brief review of the approach by Kitamoto and collaborators \cite{Kitamoto}. In Sections 4 and 5 we then demonstrate that the instability discovered in \cite{Kitamoto} is only locally measurable if a clock field is introduced.

We will work in natural units in which the speed of light, Planck's constant and the Boltzmann constant are set to one. As our cosmological background we consider a spatially flat Friedmann-Lemaitre-Robertson-Walker metric given by
\be
ds^2 \, = \, - dt^2 + a(t)^2 d{\bf x}^2 \, 
\ee
where $t$ is the background time, ${\bf x}$ are the comoving spatial coordinates, and $a(t)$ is the cosmological scale factor. The Hubble expansion parameter is given by $H(t) = {\dot a} / a$, where an overdot represents the derivative with respect to time. A key length scale for the evolution of cosmological fluctuations is the Hubble radius
\be
l_H(t) \, \equiv \, H^{-1}(t) \, .
\ee
Fluctuation modes have constant comoving wavenumber, and their physical wavelength grows as $a(t)$. Modes with a wavelength smaller than the Hubble radius oscillate while those with a larger wavelength freeze out and become squeezed (see e.g. \cite{MFB} for a review of the theory of cosmological perturbations, and \cite{RHBfluctsrev} for an overview). In accelerating backgrounds such as de Sitter space the wavelength of fluctuation modes increases faster than the Hubble radius. Hence, the phase space of {\it infrared modes} (modes with wavelength larger than the Hubble radius) increases. We will often use conformal time $\eta$ in terms of which the background metric is
\be
ds^2 \, = \, a(t(\eta))^2 [- d\eta^2 + d{\bf x}^2] \, .
\ee
The derivative with respect to conformal time will be denoted by a prime. Space-time indices are denoted by Greek letters while Latin letters run over spatial indices only. The Einstein summation convention is assumed.  A general metric is given by the metric tensor $g_{\mu \nu}$ whose determinant is denoted by $g$.

\section{Back-Reaction of Cosmological Perturbations}

In this section we review the approach to the instability of de Sitter space-time based on the back-reaction of cosmological perturbations  in the presence of matter \cite{ABM}. In early universe cosmology one considers fluctuations of the matter and metric about a homogeneous and isotropic classical background. These fluctuations are analyzed at the level of linearized perturbation equations and can be canonically quantized . At the linear level, each spatial Fourier mode of the fluctuations evolves independently. While in this approach the linear fluctuation equations are satisfied, the metric which is obtained by adding the linear fluctuations to the background does not satisfy the full Einstein equations beyond linear order. At next to leading order, a Fourier mode with wave number ${\bf k}$ can combine with a ${\bf k^{\prime}}$ mode to produce a contribution to the mode with wave number ${\bf k} + {\bf k^{\prime}}$. In particular, each ${\bf k}$ mode can combine with the ${\bf -k}$ mode to produce a contribution to the ${\bf k = 0}$ background. Thus, at second order in the perturbative expansion (the expansion parameter being the relative amplitude ${\cal{A}}$ of the metric fluctuations), the presence of fluctuations changes the background metric. This change is what is called the {\it back-reaction} effect.

Metric fluctuations can be decomposed into scalar, vector and tensor modes. The scalar modes are induced by matter fluctuations, the tensor modes correspond to gravitational waves. The vector modes are usually neglected in an expanding universe since their amplitude decays in time. In the presence of matter, the scalar modes typically have a larger effect than the tensor modes \footnote{In the context of inflation, the scalar modes are enhanced by a factor of $\epsilon^{-1}$ compared to the tensor modes, where $\epsilon \ll 1$ is the inflationary slow-roll parameter.}, and hence we focus on the scalar modes. In a de Sitter background, the phase space of {\it ultraviolet} (UV) modes (modes with a physical wavenumber $k_p$ between the Hubble value $H$ and the ultraviolet cutoff scale $\Lambda$) is constant, but the phase space of infrared (IR) modes is increasing. Note that a constant physical UV cutoff implies that comoving modes of the Hilbert space of the effective field treatment of fluctuations must be continuously created \footnote{This is the source of the {\it Trans-Planckian problem} for inflationary fluctuations \cite{Jerome} which is a motivation for the TCC \cite{TCC1, TCC2}.}. If we assume that the sub-Hubble modes are in their vacuum state, their effect can be renormalized to zero. On the other hand, the magnitude of the back-reaction effect of the infrared modes will be an increasing function of time.

The effect of cosmological perturbations on the background metric can be described by an effective energy-momentum tensor $\tau_{\mu \nu}$ of cosmological fluctuations \cite{ABM}. Here, $\tau_{\mu \nu}$ is obtained in the same way that the effective energy-momentum tensor of gravitational waves can be defined \cite{Hartle}: we consider the metric
\be
g_{\mu \nu} \, = \, g^{(0)}_{\mu \nu} + h_{\mu \nu} \, ,
\ee
where $g^{(0)}_{\mu \nu}$ is the homogeneous and isotropic background metric and $h_{\mu \nu}$ are the metric fluctuations about this background which are of the order $0 < {\cal{A}} \ll 1$. Similarly, we split the matter fields into homogeneous and isotropic matter plus matter fluctuations. We then insert this ansatz for the metric and matter into the Einstein action and expand to second order in ${\cal{A}}$. The terms linear in ${\cal{A}}$ cancel since the fluctuations are assumed to obey the linear fluctuation equations, and the background is assumed to obey the Friedmann equations. We then take all terms quadratic in the metric and matter fluctuations to the matter side of the equation, and these terms define $\tau_{\mu \nu}$. Finally, we take the spatial average of $\tau_{\mu \nu}$ to obtain the leading order effect of the fluctuations on the background (see \cite{RHBBRrev} for details).

It was shown in \cite{ABM} that $\tau_{\mu \nu}$ acts as a negative contribution to the effective cosmological constant whose magnitude grows linearly in time $t$. A heuristic reason to understand this result is that spatial gradient terms and kinetic terms are suppressed for the dominant mode of the fluctuation fields on super-Hubble scales, and that hence the equation of state of $\tau_{\mu \nu}$ \footnote{Recall that we are computing the contribution of the super-Hubble modes.} must satisfy $p = - \rho$. Since a matter overdensity creates a negative gravitational potential well, and since gravitational energy dominates over matter energy on super-Hubble scales, we obtain a negative contribution to the cosmological constant.

Unruh, however, raised the question \cite{Unruh} on how this effect could be physically measured by local observers. In the original work of \cite{ABM}, $\tau_{\mu \nu}$ was computed as a function of background time, but background time is not a physical observable since it is always possible to consider time reparametrizations. In a cosmological space-time containing perfect fluid matter with energy density $\rho$, pressure $p$ and normalized four velocity $u^{\alpha}$ with $u^{\alpha}u_{\alpha} = -1$, a physical measure of the local expansion is the divergence of the four velocity, namely
\be \label{divergence}
\Theta \, \equiv \, u^{\alpha}_{, \alpha} \, = \, \frac{1}{\sqrt{-g}} \partial_{\mu} (\sqrt{-g} u^{\mu}) \, .
\ee
In the case of a homogeneous and isotropic background metric, $\Theta = 3H$ gives the Hubble expansion rate. More precisely, one considers the spatial average of $\Theta$. It was then shown \cite{GG1} (see also \cite{AW, AB, Tanaka}) that in the case of a single matter field, the back-reaction effect is not measurable. Specifically, in the case of matter being described by a single real scalar field $\varphi$, one can show that $\Theta$ computed on a constant $\varphi$ surface is the same with and without cosmological fluctuations.

In the real world, however, we have several matter components. For instance, in late time cosmology one has cold matter ($p_m = 0$) - the dominant fluid - and radiation ($p_r = \frac{1}{3} \rho_r$) - the subdominant fluid \footnote{The dark energy fluid is irrelevant for this discussion.}. Time is measured in terms of the temperature of the radiation fluid. Since time can be re-scaled arbitrarily, e.g. by shifting the time of the Big Bang away from $t = 0$ , it does not make sense to ask questions such as ``What is the amplitude of the density fluctuation at time $t$?''. The right question to ask is ``What is the amplitude of the density fluctuations when the radiation temperature is $3^{o}K$?''. Similarly, in the early universe there are several matter fields, the matter field $\varphi$ which we take to be the dominant one and the one responsible for producing the phase of accelerated expansion, and other fields, including those which describe the Standard Model particles. For simplicity one can consider a single clock field $\chi$, a field which has a negligible effect on the curvature of space-time. In \cite{GG2} (see also \cite{conf}) it was shown, again to leading order in perturbation theory, that the back-reaction effect of infrared modes does not vanish if we compute the effect as a function of the clock field. The analysis was extended in \cite{Marozzi} to verify that the effect of back-reaction indeed leads to a dynamical relaxation of the effective cosmological constant. In \cite{Leila}, the same result was obtained in an all-order analysis in ${\cal{A}}$, but to leading order in the gradient expansion.

Recently, Kitamoto and collaborators \cite{Kitamoto, Kitamoto2} presented an analysis of the instability of de Sitter space in the presence of fluctuations. They consider only a single matter field and compute the dependence of the effective cosmological constant on time, finding logarithmic decay. At the surface, this result appears to be in contradiction with those of \cite{GG1, AW, AB}. Here, we reconsider the analysis of \cite{Kitamoto} and show that the decay which is found in that work is not locally measurable unless a separate clock field is introduced. If this is done, then the logarithmic decay of de Sitter space is recovered \footnote{Note that a number of analyses which claim to show that there is no infrared instability of de Sitter space (see e.g. \cite{Senatore}) contain only a single matter field.}.

\section{Review of the Work by Kitamoto et al.}
 
 Kitamoto et al. \cite{Kitamoto} begin with the metric written in the following form
 \be \label{metric1}
 g_{\mu \nu} \, = \, \Omega^2(x) {\tilde{g}}_{\mu \nu} \, ,
 \ee
 with
 \ba
 \Omega(x) \, &=& \, a(\eta) \phi(x) \nonumber \\
 \phi(x) \, &=& \, e^{\omega(x)} \\
 {\tilde{g}}_{\mu \nu} \, &=& \, \eta_{\mu \rho} \bigl( e^{h(x)} \bigr)^{\rho}_{\nu} \, , \nonumber
 \ea
 where $\eta_{\mu \nu}$ is the Minkowski metric and the matrix $h_{\mu \nu}$ is traceless. 

Making use of this expansion, the Einstein-Hilbert action 
\be
S \, = \, \int d^4x \sqrt{-g} \bigr[ R - 6H^2 \bigl] 
\ee
(where $R$ is the Ricci scalar of the metric $g_{\mu \nu}$ and $g$ is the determinant of the this metric) takes the following form
\be
S \, = \, \int d^4x \bigr[ \Omega^2 {\tilde{R}} + 
6 {\tilde{g}}_{\mu \nu} \partial_{\mu} \Omega \partial_{\nu} \Omega 
- 6 H^2 \Omega^2 \bigl] \, ,
\ee
where ${\tilde{R}}$ is the Ricci scalar of the metric ${\tilde{g}}_{\mu \nu}$, and we have assumed the presence of a cosmological constant which in the absence of fluctuations would induce exponential expansion with Hubble rate $H$:
\be
a(\eta) \, = \, - \frac{1}{H \eta} \, ,
\ee
recalling that in the case of exponential expansion
\be
\eta(t) \, = \, - \frac{1}{H} e^{- H t} \, .
\ee

Assuming that the back-reaction effect of fluctuations will lead to a time dependence of $H$
of the form
\be
H^2(\eta) \, = \, H^2 \bigr( \frac{1}{-H\eta} \bigl)^{- 2 \gamma} \, ,
\ee
where $\gamma$ is a constant to be determined, the authors of \cite{Kitamoto} make the following ansatz for the action
\be
S_{*} \, = \, \int d^4x \sqrt{-g} \bigr[ R - 6H^2(\gamma) e^{- 2\Gamma(\gamma) \varphi}  
- 2 \Gamma(\gamma) g^{\mu \nu} \partial_{\mu} \varphi \partial_{\nu} \varphi \bigl] 
\ee
where $\varphi$ is a scalar field which represents the conformal factor of the metric fluctuations, and $\Gamma$ is another constant which depends on $\gamma$. In terms of the conformally related metric ${\tilde{g}}_{\mu \nu}$ we obtain
\ba
S_{*} \, &=& \, \int d^4x \bigr[ \Omega^2 {\tilde{R}} - 6H^2(\gamma) \Omega^{4( 1 - \Gamma(\gamma)/2)} \\
& & + \bigr( 6 - 2 \Gamma(\gamma) \bigl) {\tilde{g}}^{\mu \nu} \partial_{\mu} \Omega \partial_{\nu} \Omega \bigl] \, . \nonumber
\ea

The main result of \cite{Kitamoto} is that (in the absence of extra matter fields) the infrared fluctuations of the conformal mode lead to a screening of the Hubble expansion rate
\be \label{decay1}
H(t)^2 \, = \, H^2 \bigr( 1 - 2 \gamma {\rm ln}(a(t)/a(t_i)) \bigl) \, ,
\ee
where $H$ is the Hubble expansion rate in the absence of fluctuations, and $t_i$ is the initial time, the time at which $H(t) = H$. This represents the screening of the cosmological constant. Note that the time dependence of the screening is the same found originally in \cite{ABM} and, in the context of the corpuscular theory of gravity, in \cite{Dvali}.
 
\section{Physical Measurability}
  
We will now demonstrate that the decay given by (\ref{decay1}) is not locally physically measurable unless a clock field is introduced. Recall that the local physical observable $\Theta$ describing the expansion rate of space is given in (\ref{divergence}).  In order to compute this variable we need to solve the geodesic equation for the four velocity $u^{\alpha}$
\be
\frac{d u^{\mu}}{d \tau} + \Gamma^{\mu}_{\alpha \beta} u^{\alpha} u^{\beta} \, = \, 0 \, ,
\ee
where the $\Gamma^{\mu}_{\alpha \beta}$ are the Christoffel symbols of the metric $g_{\mu \nu}$. For the unperturbed space-time, the four velocity is given by
\be
u^{\alpha} \, = \, \bigr( 1, 0, 0, 0 \bigl) \, ,
\ee
and, in the presence of fluctuations, $u^{\alpha}$ can be expanded to second order
\be
u^{\alpha} \, = \, 
\bigr( 1 + \delta^{(1)} u^0 + \delta^{(2)} u^0, \delta^{(1)} u^i + \delta^{(2)} u^i \bigl) \, ,
\ee
where the superscripts $(1)$ and $(2)$ indicate the order of the terms in the perturbative expansion.

To evaluate $u^{\alpha}$, we start with the metric (\ref{metric1}) which we can write in compact form as
\be
{\bf{g}} \, = \, a^2 e^{2 \omega} e^{{\bf{h}}} {\bf{\eta}} \, ,
\ee
where the bold face symbols stand for matrices, and $\omega$ is a function. Both ${\bf{h}}$ and $\omega$ are linear in the fluctuation amplitude. Expanding to second order, the metric and its inverse then become
\ba
g_{\mu \nu} \, &=& \, a^2 \bigr( \eta_{\mu \nu} + f_{\mu \nu} \bigl) \\
g^{\mu \nu} \, &=& \, a^{-2} \bigr( \eta^{\mu \nu} + p^{\mu \nu}  \bigl) \nonumber 
\ea
with
\ba
f_{\mu \nu} \, &=& \, 2\omega\eta_{\mu \nu} + h_{\mu \nu} + 2 \omega^2 \eta_{\mu \nu}
+ 2 \omega h_{\mu \nu} + \frac{1}{2} h_{\mu}^{\rho} h_{\rho \nu} \\
p_{\mu \nu} \, &=& \, - 2\omega\eta_{\mu \nu} - h_{\mu \nu} + 2 \omega^2 \eta_{\mu \nu}
+ 2 \omega h_{\mu \nu} + \frac{1}{2} h_{\mu}^{\rho} h_{\rho \nu} \, . \nonumber
\ea
Note that $\phi$ and the clock field introduced in the next section couple differently to the Einstein frame metric.

By solving the geodesic equation to second order we find
\ba \label{flow}
u^{0} \, &=& \, 1 + \frac{1}{2} f_{00} + \frac{1}{2} f^{00}f_{00} + \frac{1}{2} f^{0i}f_{0i}  \nonumber \\
u^{i} \, &=& \, - a^{-1} f_{0i} + \delta^{(2)} u^i \, ,
\ea
where $\delta^{(2)}u^i$ vanishes in the infrared limit in which we neglect all terms involving spatial derivatives. In this limit, the local expansion rate reduces to
\be \label{expansion2}
\Theta \, = \, \frac{1}{\sqrt{-g}} \partial_0 (\sqrt{-g}) u^0 + \partial_0 u^0 \, .
\ee

We can always make a gauge choice such that $h_{00} = 0$ (partial synchronous gauge). There is, however, still a second gauge degree of freedom. We choose this to make the computation of the average expansion rate easy for the spatial hypersurface we choose. In the absence of a separate clock field, the canonical choice is to consider the hypersurfaces of constant scalar field $\varphi$. If we choose our time variable such that constant time corresponds to constant $\varphi$, this implies  considering a constant value of $\omega$, and since $\omega$ has to average to zero over the constant, we have $\omega = 0$. Hence, $u^0 = 1$.

Our goal is to compute the spatial average of the local expansion rate. Without an additional clock field, the only physically well-defined averaging surface is the surface of constant $\varphi$. In general, the spatial average of an observable $A$ is defined by
\be \label{average}
<A> \, = \, \frac{\int d^3x \sqrt{\gamma} A}{\int d^3x \sqrt{\gamma} } \, 
\ee
where here $\gamma$ is the determinant of the induced metric $\gamma_{\mu \nu}$ on the constant $\varphi$ surface which we choose. As we have shown in the previous paragraph, for our choice of the surface $u^{0} = 1$ is constant. Since $h$ is traceless, the determinant of the metric $\gamma$ for our surface and in the gauge we have chosen is the same as the determinant of the background metric, and hence it immediately follows that
\be
\Theta \, = \, 3H \, ,
\ee
and we reach the conclusion that, in the absence of an independent clock field, the decay of $H$ found in (\ref{decay1}) is not physically measurable.

\section{Decay in the Presence of a Clock Field}

In the previous section we have seen that the instability in the measured Hubble expansion rate is not physically measurable if the matter sector contains only one field. Let us now introduce a clock field $\chi$. We consider the clock field to be perturbed in the coordinate system in which the temporal coordinate is set by dominant field $\varphi$. Equivalently, using coordinates in which time is set by the value of the clock field, $\varphi$ (and thus $\omega$) are fluctuating. In this case, in the presence of fluctuations of the dominant matter field, the clock field will not be comoving with the dominant matter.

We are thus interested in computing the spatially averaged expansion rate
\be
H_{eff} \, \equiv \, \frac{1}{3} < \Theta > \, = \, 
\frac{1}{3} \frac{\int d^3x \sqrt{\gamma} \Theta}{ \int d^3x \sqrt{\gamma}} \, ,
\ee
where the integral is over the constant $\chi$ hypersurfaces. Specifically, we are interested in computing the effect of super-Hubble mode fluctuations on the local expansion rate. For these modes, we can neglect all terms containing spatial derivatives and thus work in the leading order gradient expansion \footnote{Note that there is an effect of spatial gradient modes, as computed in \cite{Notari, Lam}. However, this effect acts like a contribution to the spatial curvature, and not like a contribution to the effective cosmological constant.}.

In \cite{Leila}, an expression for $H_{eff}$ in terms of the longitudinal gauge fluctuations $\phi(x, t)$ and $\psi(x, t)$ in which the metric takes the form
\be
g_{\mu \nu} \, = \, a^2 {\rm{diag}} ( - e^{2\phi}, e^{2\psi}, e^{2\psi}, e^{2\psi} ) \, ,
\ee
was derived in the leading order gradient expansion, i.e. neglecting all terms with spatial gradients. To leading order in the amplitude of fluctuations, the result for the difference $\Delta H_{eff}$ between the actual effective Hubble expansion rate and the expansion rate in the absence of fluctuations is
\be
\Delta H_{eff} \, = \, 3 < \psi {\dot{\psi}} > \, ,
\ee
where the angular brackets indicate the spatial average over the constant longitudinal gauge time slices \footnote{This result was based on earlier work of \cite{Venturi}.}.

In the case of the conformal mode considered here, the metric in longitudinal gauge is
\ba
g_{00} \, &=& \, - a^2 e^{2 \omega} \nonumber \\
g_{0i} \, &=& \, 0 \\
g_{ij} \, &=& \, a^2 e^{2 \omega}  \delta{ij} \nonumber
\ea
and hence 
\be \label{result}
\Delta H_{eff} \, = \, 3 < \omega {\dot{\omega}} > \, .
\ee
 
At linear order in the fluctuations, each Fourier mode ${\tilde{\omega}}_k$ of $\omega$ evolves independently. On the super-Hubble scales whose back-reaction we wish to study, the dominant mode of ${\tilde{\omega}}$ is constant, and the second mode is decaying \footnote{This is a well-known result from the theory of cosmological perturbations (see e.g. \cite{Mukh, MFB}).}:
\be \label{scaling}
{\tilde{\omega}}_k(t) \, = \, \omega^{(1)}_k + \Omega^{(2)}_k e^{-H(t - t_H(k))} \, ,
\ee
where $t_H(k)$ is the time when the mode $k$ exits the Hubble radius (and the formula is valid when the scale is super-Hubble, i.e. when $t > t_H(k)$), and the amplitudes $\omega^{(1)}$ and $\omega^{(2)}$ are independent of time. Since the mode functions are oscillating on sub-Hubble scales, we expect that the amplitudes of the two modes at Hubble radius crossing will have the same magnitude, and, in particular, they have the same sign (which without loss of generality we can take to be positive). With vacuum initial conditions for the fluctuations $\omega_k$ we hence have
\be \label{mode}
\omega^{(1)} \, \sim \, \omega^{(2)} \, \sim \, {\cal{A}} k^{-3/2} \, ,
\ee
where ${\cal{A}}$ is the amplitude of the quantum vacuum fluctuations
\be
{\cal{A}} \, \sim \, \frac{H}{m_{pl}} \, ,
\ee
$m_{pl}$ being the Planck mass.
 
To evalute the correction term (\ref{result}) to the local expansion rate, we make use of Parceval's theorem
\be \label{Parceval}
\int d^3x \omega(x) {\dot{\omega}}(x) \, = 
\, \int d^3k {\tilde{\omega}}_k {\dot{\tilde{\omega}}}_k \, .
\ee
where we use the following convention for the Fourier transform
\be 
\omega(x) \, = \, V^{1/2} (2 \pi)^{-3/2} \int d^3k {\tilde{\omega}}_k e^{i k x} \, ,
\ee
($V$ is the spatial volume introduced as a cutoff). Thus, we see that each Fourier mode of the linear fluctuations contributes to the average. As argued in Section II, in an exponentially expanding background, the contribution of sub-Hubble modes is time-independent since the phase space of these UV modes does not increase. On the other hand, the phase space of super-Hubble modes increases since modes continue to exit the Hubble radius and hence the phase space of these modes is building up.

Making use of (\ref{Parceval}), it then follows from (\ref{result}) that the contribution of the infrared modes to the change in the local expansion rate averaged over the constant clock time hypersurfaces is
\be \label{result2}
\Delta H_{eff} \, = \, 3 \int_{k_{IR}}^{k_H(t)}d^3k \omega_k {\dot{\omega}}_k  \, .
\ee
where $k_{IR}$ is a physical infrared cutoff which can be taken to be the comoving $k$ of the initial Hubble radius, and $k_H(t)$ is the wavenumber corresponding to Hubble radius crossing at time $t$, i.e.
\be
k_H(t) \, = \, a(t) H \, 
\ee
(neglecting the time dependence of $H$). Inserting the time dependence (\ref{scaling}) and amplitude (\ref{mode}) of the mode functions, we obtain
\be \label{result3}
\Delta H_{eff} \, = \, - 12 \pi {\cal{A}}^2 H \int_{k_{IR}}^{k_H(t)}dk k^{-1} e^{-H(t - t_H(k))}  \, ,
\ee
where $t_H(k)$ is the time when the mode $k$ exits the Hubble radius.  

The first conclusion which can be drawn from the above result (\ref{result3}) is that the correction to the Hubble expansion rate is negative. This is in agreement with previous works, and in particular with the work of \cite{Kitamoto} which find that the leading contribution of infrared modes leads to a decrease in the local expansion rate. In fact, using the fact that $e^{Ht_H(k)} = k/H$ we can write (\ref{result3}) as \cite{Leila}
\be \label{result4}
\Delta H_{eff} \, = \, - 12 \pi {\cal{A}}^2 H  \bigr( 1 - \frac{k_{IR}}{H} e^{-H(t - t_i)} \bigl) \, ,
\ee
where $t_i$ is the initial time (the time when the scale $k_i$ exits the Hubble radius). From this result we immediately see the magnitude of the back-reaction effect increases as the phase space of infrared modes increases, i.e. 
\be \label{result4}
{\dot{\Delta H_{eff}}} \, < \, 0 \, .
\ee
In \cite{Leila}, the above analysis has been generalized to arbitrary order in perturbation theory (the perturbative expansion as a function of ${\cal{A}}$, but still to leading order in the gradient expansion. The same conclusion will be valid for us.

The above analysis is based on neglecting the time-dependence of $H$ in the computation of the fluctuations. The time dependence of $H$ which is induced by the back-reaction which we study will yield correction terms to the ones we have computed, but these terms will be of higher order in perturbation theory. Note that, in the spirit of considering a clock field, we are neglecting the effects of $\chi$ on the expansion rate.
 
\section{Conclusions and Discussion} \label{conclusion}

 We have studied the approach to the relaxation of a bare cosmological constant proposed in \cite{Kitamoto}. The instability discussed in \cite{Kitamoto} is due to infrared fluctuations. We have shown that the decay computed in \cite{Kitamoto} is not physically measurable using the spatial average of the local expansion rate, unless a separate clock field is introduced. This agrees with the expectation from previous studies which indicate that effects of super-Hubble modes are not locally measurable if the fluctuations are purely adiabatic and no separate clock field is introduced.
 
In the absence of a separate clock field, the spatial averaging must be done over surfaces of constant value of the single matter field which is present. In this case, the back-reaction term studied in \cite{Kitamoto} has no locally measurable effect. However, in the presence of a clock field, the spatial averaging must be done over a different surface, namely a surface of constant value of the clock field. We have shown that in this case the instability discussed in \cite{Kitamoto} is recovered: the local expansion rate decreases. The correction factor is quadratic in the amplitude of the linear fluctuations, and it increases as the phase space of infrared modes builds up.

Our analysis only yields the contributions to the local expansion rate which are leading order in conventional perturbation theory. It would be of interest to complement the analysis we have done here with a non-perturbative approach, keeping track of the spatial gradient terms. Work on this topic is in progress.    

\section*{Acknowledgement}

\noindent The research at McGill is supported in part by funds from NSERC and from the Canada Research Chair program. RB is grateful for hospitality of the Institute for Theoretical Physics and the Institute for Particle Physics and Astrophysics of the ETH Zurich. VC is supported in part by a fellowship from the FQRNT. LH is supported by funding from the European Research Council (ERC) under the European
Unions Horizon 2020 research and innovation programme grant agreement No 801781 and
by the Swiss National Science Foundation grant 179740.

\end{document}